\documentclass[aps,showpacs,preprintnumbers,amsmath, amssymb]{revtex4}

\usepackage{float}
\usepackage{graphics,epsfig}
\usepackage{graphicx}
\usepackage{dcolumn}
\usepackage{bm}

\begin{document}
\baselineskip=0.8 cm

\title{{\bf Dark energy model interacting with dark matter and unparticle}}
\author{Songbai Chen}
\email{csb3752@163.com} \affiliation{ Institute of Physics and
Department of Physics, Hunan Normal University,  Changsha, Hunan
410081, P. R. China \\ Key Laboratory of Low Dimensional Quantum
Structures \\ and Quantum Control of Ministry of Education, Hunan
Normal University, Changsha, Hunan 410081, P. R. China}

\author{Jiliang Jing}
\email{jljing@hunnu.edu.cn}
 \affiliation{ Institute of Physics and
Department of Physics, Hunan Normal University,  Changsha, Hunan
410081, P. R. China \\ Key Laboratory of Low Dimensional Quantum
Structures \\ and Quantum Control of Ministry of Education, Hunan
Normal University, Changsha, Hunan 410081, P. R. China}

\vspace*{0.2cm}
\begin{abstract}
\baselineskip=0.6 cm
\begin{center}
{\bf Abstract}
\end{center}

We study dynamical behaviors of the dark energy models interacting
with dark matter and unparticle in the standard flat FRW cosmology.
We considered four different interacting models and examined the
stability of the critical points. We find that there exist late-time
scaling attractors corresponding to an accelerating Universe and the
alleviation of the coincidence problem depends on the choice of
parameters in the models.
\end{abstract}

 \maketitle
\newpage
\section{Introduction}

Many observations confirm that we live in an accelerating Universe.
Within the framework of Einstein gravity, this acceleration can be
explained by the mysterious energy component named dark energy (DE)
with negative pressure, which occupies almost $70\%$ of the content
of the universe at present \cite{A1,A2,A3,A4}. The simplest
explanation for dark energy is the cosmological constant, which is a
term that can be added to Einstein's equations. This term acts like
a perfect fluid with an equation of state (EoS) $\omega_x=-1$, and
the energy density is associated with quantum vacuum.  Although this
interpretation is consistent with observational data, at the
fundamental level it fails to be convincing. The vacuum energy
density is far below the value predicted by any sensible quantum
field theory, and it suffers the coincidence problem, namely,  why
the DE and the Dark Matter (DM) are comparable in size exactly right
now. To overcome the coincidence problem, some sophisticated
dynamical scalar field DE models, such as quintessence, k-essence
and phantom field, have been put forth to replace the cosmological
constant \cite{T1}. However, these scalar field DE models can not
still resolve the coincidence problem.

The interacting DE model has been regarded as a possible way to
alleviate this problem \cite{c1}. In these models it is assumed that
there exists a nonzero interaction between DE and DM in the Universe
that gauges DE transfers to DM which allows us to create an
equilibrium balance in the evolution of the Universe, so that the
density of DE keeps the same order as that of DM at late times.
Although lacking of a fundamental explanation for the source of the
interaction, the interacting models are manifested as the useful and
robust models at the order of one standard deviation by some data
sets from observational cosmology including CMB shift parameter,
BAO, age parameter and supernova observations and so on
\cite{c1,hb,c4}. Therefore, the interacting models have attracted a
great deal of interest \cite{c2,c3,c5,c6,c7,c8,pt,dM1,dE1,pq,
BM1,yg1,sb1,phw1,phw2,gf,gf1}. Recently, some attempts have been
proposed to describe the interaction between DE and DM from a
fundamental field theory point of view \cite{tf1,tf2}.

The investigations above have been focused only on the interaction
between DE and DM. However, it is physically reasonable and even
expected from a theoretical point of view, that DM as well as DE can
interact with other dark components of the universe \cite{tw1,tw2}.
Recent investigations shows that unparticle can be treated
theoretically as another important dark components in the Universe.
The concept of unparticle is introduced first by Georgi \cite{u1},
which is based on the hypothesis that there could be an exact scale
invariant hidden sector resisted at a high energy scale (for a
recent review of unparticles, see \cite{u2,u3}). Although the
fundamental energy scale of such a sector is far beyond the reach of
today¡¯s or near future accelerators, it is possible that this new
sector affects low energy phenomenology. These effects are described
as unparticle in the effective low energy field theory  because that
the behaviors of these new degrees of freedom are quite a different
those of the ordinary particles. For example, their scaling
dimension does not have to be an integer or half an integer.
Recently, a lot of work \cite{u5,u51} have been focused on the new
collider signals for unparticle physics. One of interesting feature
of unparticle is that it does not have a definite mass and instead
has a continuous spectral density as a consequence of scale
invariance \cite{u1}
\begin{eqnarray}
\rho(P^2)=A_{d_u}\theta(P^0)\theta(P^2)(P^2)^{d_u-2},
\end{eqnarray}
where $P$ is the 4-momentum, $A_{d_u}$ is the normalization factor
and $d_u$ is the scaling dimension. The theoretical bounds of the
scaling dimension $d_u$ are $1\leq d_u\leq 2$ (for boson unparticle)
or $3/2\leq d_u\leq 5/2$ (for fermion unparticle) \cite{u3}. The
pressure and energy density of the thermal boson unparticle are
given by \cite{u4}
\begin{eqnarray}
p_u&=&g_sT^4\bigg(\frac{T}{\Lambda_u}\bigg)^{2(d_u-1)}\frac{\mathcal{C}(d_u)}{4\pi^2},\nonumber\\
\rho_u&=&(2d_u+1)g_sT^4\bigg(\frac{T}{\Lambda_u}\bigg)^{2(d_u-1)}\frac{\mathcal{C}(d_u)}{4\pi^2},
\end{eqnarray}
where $\mathcal{C}(d_u)=B(3/2,d_u)\Gamma(2du+2)\zeta(2du + 2)$,
while $B$, $\Gamma$, $\zeta$ are the Beta, Gamma and Zeta functions,
respectively. Thus, the EoS of boson unparticle reads \cite{u4}
\begin{eqnarray}
\omega_u=\frac{1}{2d_u+1}.\label{wu}
\end{eqnarray}
For the fermion unparticle, we find the EoS has the same form as
that of boson one. Obviously, the EoS of unparticle $\omega_u$ is
positive which is different from that of DE and DM. Since the
unparticle interacts weakly with standard model particles, it can be
regarded as a new form of dark component. Recent investigations show
that the unparticles  play an important role in the early universe
\cite{u6} and black hole physics \cite{u7}. Therefore, it is natural
to ask whether there exists some new properties in the late-time
evolution of the Universe if the unparticle takes part in the
interaction with DE and DM?  In this paper we choose four different
coupling terms and study the dynamical behaviors of the interacting
DE models with DM and unparticle by the phase space analysis method
to discuss further stability of the critical points and their
cosmological implications.

The paper is organized as follows: in sections II, we construct a
cosmological scenario in which DE interacts with DM and unparticle,
and then we present the formalism for its transformation into an
autonomous dynamical system which is suitable for a phase space
stability analysis. In Sec.III, we consider some special coupling
forms and perform the phase space analysis of the corresponding
interacting DE model with DM and unparticle, and then discuss their
cosmological implications. Our conclusions and discussions will be
presented in the last section.

\section{Interacting dark energy model with dark matter and unparticle}

In the Einstein theory, the a flat FRW universe is described by the
standard Friedmann equation and Raychaudhuri field equation
\begin{eqnarray}
&&H^2=\frac{\kappa}{3}\rho,\label{E1}\\
&&\dot{H}=-\frac{\kappa}{2}(\rho+p).\label{E2}
\end{eqnarray}
$H$ is the Hubble parameter and $\kappa$ is the constant $8\pi G$.
The total energy density $\rho=\rho_m+\rho_x+ \rho_u$, where
$\rho_m$, $\rho_x$ and $\rho_u$ correspond to the energy densities
of DM,  DE and unparticle, respectively. For simplicity here we have
neglected the radiation and baryons since we are concentrating on
the late time accelerating Universe.

The interaction among DE, DM and unparticle can be described in the
background by the balance equations
\begin{eqnarray}
&&\dot{\rho_x}+3H(1+\omega_x)\rho_x=\Gamma_1,\nonumber\\
&&\dot{\rho_m}+3H\rho_m=\Gamma_2,\nonumber\\
&&\dot{\rho_u}+3H(1+\omega_u)\rho_u=\Gamma_3 \label{2}
\end{eqnarray}
Here the terms $\Gamma_1$, $\Gamma_2$ and $\Gamma_3$ describe the
coupling among DE, DM and unparticles. The total conservation
equation demands that
\begin{eqnarray}
\Gamma_1+\Gamma_2+\Gamma_3=0.
\end{eqnarray}
To analyze the evolution of the dynamical system, we introduce the
dimensionless variables
\begin{eqnarray}
x\equiv\frac{\kappa\rho_x}{3H^2},\;\;\;\;\;\;y\equiv\frac{\kappa\rho_m}{3H^2},\;\;\;
z\equiv\frac{\kappa\rho_u}{3H^2},\;\;\;\;\;\;\;
\frac{d}{d\;N}=\frac{1}{H}\frac{d}{d\;t},\label{E4}
\end{eqnarray}
where $N\equiv\ln{a}$ is the number of $e$-folding to represent the
cosmological time. Using the above definitions, the Hubble equations
can be rewritten as
\begin{eqnarray}
x+y+z=1,
\end{eqnarray}
and
\begin{eqnarray}
\frac{\dot{H}}{H^2}=-\frac{3}{2}\bigg[1+\frac{\omega_xx+\omega_uz}{x+y+z}\bigg]=-\frac{3}{2}(1+\omega_xx+\omega_uz).\label{EHH}
\end{eqnarray}
The effective total EOS $\omega_{tot}$ is given by
\begin{eqnarray}
\omega_{tot}=\frac{\omega_x\rho_x+\omega_u\rho_u}{\rho_x+\rho_m+\rho_u}=\frac{\omega_xx+\omega_uz}{x+y+z}=\omega_xx+\omega_uz.\label{st1}
\end{eqnarray}
Once the concrete forms of the coupling terms $\Gamma_1$, $\Gamma_2$
and $\Gamma_3$ are given, the equations of motion (\ref{E1}),
(\ref{E2}) and (\ref{2}) can be transformed to an autonomous system
containing the variables $x$ and $y$ and their derivatives with
respect to $N =\ln a$. This autonomous system has generally the
form: $X'=f(X)$, where $X$ is the column vector constituted by the
auxiliary variables, $f(X)$ is the corresponding column vector of
the autonomous equations, and prime denotes derivative with respect
to $N =\ln a$. Solving the equations $X'=0$, we can obtain the
critical points $X_c$. Then, in order to describe the stability
properties of these critical points, we can expand the equations
$X'=f(X)$ around $X_c$ by setting $X=X_c +\delta U$ with $\delta U$
the perturbations of the variables considered as a column vector.
Thus, for each critical point we can expand the equations for the
perturbations up to the first order as: $\delta U' = \Xi \cdot
\delta U$, where the matrix $\Xi$ contains the coefficients of the
perturbation equations. Through the analysis of the eigenvalues of
$\Xi$, we can describe the stability of each critical point. In
general, for an arbitrary coupling terms $\Gamma_i$ it is difficult
to obtain the analytical forms of the critical points. Here we only
consider some specific forms of $\Gamma_i$ and examine the stability
of interacting DE models with DM and unparticle in the next section.

\section{Analysis of stability in the phase space}

In this section, we will consider some concrete coupling forms
$\Gamma_i$ and analysis the stability of the corresponding
interacting DE models with DM and unparticle, and then discuss their
cosmological implications.

\subsection{Interacting model I }

In general, there is as yet no basis in fundamental theory for a
specific coupling in the dark sectors because that the nature of
dark sectors remain unknown. Thus, all coupling models discussed at
the present moment are necessarily phenomenological \cite{c1}. There
are two criterions to determine whether some models can be more
physical justification than the others. One is to confront
observations. The other is to examine whether the coupling can lead
to accelerated scaling attractor solutions\cite{pt}, which is a
decisive way to achieve similar energy densities in dark sectors and
alleviate the coincidence problem. Motivated by analogy with
dissipation of cosmological fluids, the coupling terms $\Gamma_i$
which are proportioned to the densities $\rho_x$ and $\rho_m$ have
been studied in the context of quintessence \cite{c1} and phantom
\cite{c7,dE1,dM1,phw1} models. In this section, we first consider
the case in which all of the coupling terms $\Gamma_i$ are
proportioned to the density of DE $\rho_x$.  Here we choose
$\Gamma_2=\Gamma_3$ so that we can obtain the analytical expression
for the critical point, which is very convenient for us to study the
dynamics of the model in the following calculations. The concrete
expressions of $\Gamma_i$ are
\begin{eqnarray}
\Gamma_1=-6bH\rho_x,\;\;\;\;\;
\Gamma_2=\Gamma_3=3bH\rho_x,\label{cou1}
\end{eqnarray}
where $b$ is a positive coupling constant. It means that in this
model only DE can convert to DM and unparticle. Using dimensionless
variables, the dynamical equations of the system can be expressed as
\begin{eqnarray}
x'&=&-6bx-3\omega_xx+3x(\omega_xx + \omega_uz),\nonumber\\
y'&=&3bx+3y(\omega_xx + \omega_uz),\nonumber\\
z'&=&3bx-3\omega_uz+3z(\omega_xx + \omega_uz).
\end{eqnarray}
Solving the equations $x'=0$, $y'=0$ and $z'=0$, we obtain the
critical points $(x_c,\; y_c,\;z_c)$:
\begin{eqnarray}
&&\cdot \;\;\text{Point}\; A_1: \;( 0,\;\;1,\;\;0),\nonumber \\ &&
\cdot \;\;\text{Point}\; B_1: \; ( 0,\;\;0,\;\;1),\nonumber \\
&&\cdot \;\;\text{Point}\; C_1: \bigg(
\frac{(2b+\omega_x)(2b+\omega_x-\omega_u)}{(\omega_x-\omega_u)(b+\omega_x)+b\omega_x},\;\;
\frac{-b(2b+\omega_x-\omega_u)}{(\omega_x-\omega_u)(b+\omega_x)+b\omega_x},\;\;\frac{-b(2b+\omega_x)}{(\omega_x-\omega_u)(b+\omega_x)+b\omega_x}\bigg).
\end{eqnarray}
The points $A_1$ and $B_1$ imply that our Universe is dominated by
DM and unparticle, respectively. The condition for the existence of
the point $C_1$ is $0<b<-\frac{\omega_x}{2}$.  The critical point
$C_1$ denotes that the DE, DM and unparticle can be coexisted in the
late-times of the Universe.

After some operations, we can obtain the $3\times 3$ matrix $\Xi$ of
the linearized perturbation equations
\begin{eqnarray}
\Xi=\bigg[ \begin{array}{ccc} \omega_x(6x_c-3)+3\omega_uz_c-6b\;\;&0&3\omega_ux_c \\
3(\omega_xy_c+b) &3(\omega_xx_c+\omega_uz_c)& 3\omega_uy_c \\
3(\omega_xz_c+b) &0 & 3\omega_x x_c+\omega_u(6z_c-3)
\end{array}\bigg].
\end{eqnarray}
The eigenvalues of the coefficient matrix $\Xi$ encode the behavior
of the dynamical system near the critical points. If the real parts
of all of the eigenvalues of the matrix $\Xi$ are negative, then the
critical point is a stable point, otherwise it is unstable. Through
some careful calculations, we obtain the eigenvalues of the
coefficient matrix for these critical points
\begin{eqnarray}
&&\cdot \;\;\text{Point} \;A_1: \;\;\;\;\; \lambda_1=0,\;\;\;\;\;
\lambda_2=-3\omega_u,\;\;\;\;\;
\lambda_3=-3(2b+\omega_x),\nonumber\\
&&\cdot \;\;\text{Point}\; B_1: \;\;\;\;\;
\lambda_1=3\omega_u,\;\;\;\;\; \lambda_2=3\omega_u,\;\;\;\;\;
\lambda_3=-3(2b+\omega_x-\omega_u),\nonumber\\
&&\cdot \;\;\text{Point}\; C_1: \;\;\;\;\;
\lambda_1=3(2b+\omega_x),\;\;\;\;\;
\lambda_2=3(2b+\omega_x),\;\;\;\;\;
\lambda_3=3(2b+\omega_x-\omega_u).
\end{eqnarray}
For the point $A_1$, the eigenvalue $\lambda_1$ is non-negative,
which indicates that $A_1$ is not a stable point. For the point
$B_1$, we also find that both of the eigenvalues $\lambda_1$ and
$\lambda_2$ are positive because that the EoS of the unparticle
$\omega_u>0$. Therefore the point $B_1$ is not the stable point. For
the point $C_1$, when $0<b<-\frac{\omega_x}{2}$, all of the
eigenvalues $\lambda_1$, $\lambda_2$ and $\lambda_3$ are negative,
which means that $C_1$ is a stable point as shown in figure (1). The
stable region of $C_1$ is not affected by the EoS of the unparticle,
but the position of $C_1$ in the phase space is decided together by
$\omega_x$, $\omega_u$ and the coupling constant $b$. From
Eq.(\ref{st1}), we also learn that the effective total EoS at point
$C_1$ is $\omega_{tot}=2b+\omega_x$, which is independent of
$\omega_u$. When $b\rightarrow-\omega_x/2$, we find that
$\omega_{tot}$ tends to zero. It is consistent with the result shown
in figure (\ref{1f}) in which as the coupling is strong enough the
Universe will be dominated by DM. Moreover, for the critical point
$C_1$, we find that in its stable region  none of the coordinates
$x_c$, $y_c$ and $z_c$ in phase space vanishes, which means that the
coincidence problem can be alleviated in the Universe described by
the point $C_1$. Since the effective total EoS at point $C_1$ is
$\omega_{tot}=2b+\omega_x$ we else obtain
$\ddot{a}\propto-(3\omega_x+6b+1)\;t^{\frac{2}{3(\omega_x+2b+1)}-2}$
and $\rho\propto a^{-3(\omega_x+2b+1)}$. It means that point $C_1$
is an accelerated scaling solution as
$b<-\frac{\omega_x}{2}-\frac{2}{3}$ and there is singularity in the
finite future as $b<\frac{\omega_x}{2}-\frac{1}{2}$.
\begin{figure}[ht]
\begin{center}
\includegraphics[width=6cm]{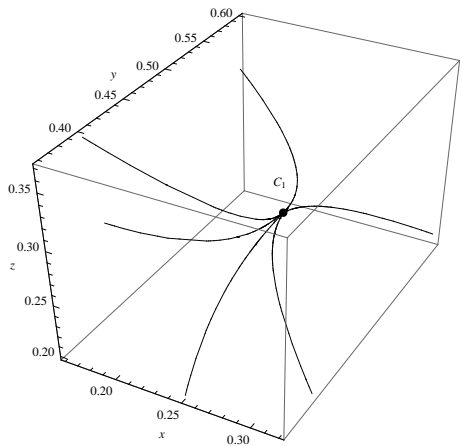}
\caption{The phase diagram of interacting dark energy with DM and
unparticle through the coupling terms (\ref{cou1}). The point $C_1$
is the critical point. Here we choose the values $\omega_x=-1.2$,
$\omega_u=0.28$ and $b=0.5$ in the stable region $b<-\omega_x/2$.}
\end{center}
\end{figure}
\begin{figure}[ht]
\begin{center}
\includegraphics[width=6cm]{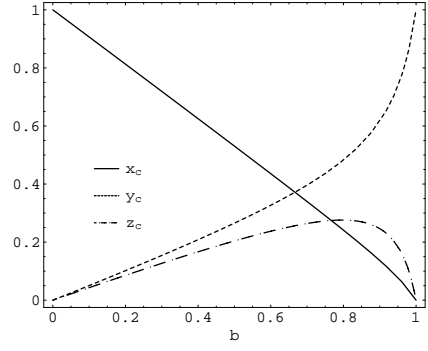}
\caption{Variety of $x_c$, $y_c$ and $z_c$ with $b$ at the critical
point $C_1$ for fixed $\omega_x=-2$ and $\omega_u=0.3$. The coupling
constant $b$ is located in the stable region $0<b<-\omega_x/2$. }
\label{1f}
\end{center}
\end{figure}

\subsection{Interacting model II }

In the previous discussions, we studied only the case DE converts to
DM and unparticle and do not consider the exchange of energy between
DM and unparticle. In this section, we will consider another special
case in which besides DE transfers to DM, DM can also be converted
to unparticle. Similarly, in order to obtain the analytical forms
for the critical points, we assume that the coupling terms
$\Gamma_i$ have the forms
\begin{eqnarray}
\Gamma_1=-3bH\rho_x,\;\;\;\; \;
\Gamma_2=3bH(\rho_x-\rho_m),\;\;\;\;\;\;
\Gamma_3=3bH\rho_m,\label{cou2}
\end{eqnarray}
respectively.

The dynamics of the system can be described by
\begin{eqnarray}
x'&=&-3bx-3\omega_xx+3x(\omega_xx + \omega_uz),\nonumber\\
y'&=&3b(x-y)+3y(\omega_xx + \omega_uz),\nonumber\\
z'&=&3by-3\omega_uz+3z(\omega_xx + \omega_uz),
\end{eqnarray}
and the critical points $(x_c,\; y_c,\;z_c)$ are
\begin{eqnarray}
&&\cdot \;\;\text{Point}\; A_2: \;( 0,\;\;0,\;\; 1),\nonumber \\ &&
\cdot \;\;\text{Point}\; B_2: \; ( 0,\;\;1-\frac{b}{\omega_u},\;\;\frac{b}{\omega_u}),\nonumber \\
&&\cdot \;\;\text{Point}\; C_2: \; \bigg(
\frac{\omega_x(b+\omega_x-\omega_u)}{\omega^2_x+\omega_u(b-\omega_x)},\;\;
\frac{-b(b+\omega_x-\omega_u)}{\omega^2_x+\omega_u(b-\omega_x)},\;\;\frac{b^2}{\omega^2_x+\omega_u(b-\omega_x)}\bigg).
\end{eqnarray}
The condition that the point $C_2$ exists is
$0<b<\omega_u-\omega_x$. Similarly, the $3\times 3$ matrix $\Xi$ of
the linearized perturbation equations reads
\begin{eqnarray}
\Xi=\bigg[ \begin{array}{ccc} -3(\omega_x+b-2\omega_xx_c-\omega_uz_c)\;\;&0&3\omega_ux_c \\
3\omega_x(y_c+b) &3(\omega_xx_c+\omega_uz_c)-3b& 3\omega_uy_c \\
3\omega_xz_c &3b &3\omega_x x_c+\omega_u(6z_c-3)
\end{array}\bigg].
\end{eqnarray}
The eigenvalues of the coefficient matrix for these critical points
are
\begin{eqnarray}
&&\cdot \;\;\text{Point} \;A_2: \;\;\;\;\;
\lambda_1=3\omega_u,\;\;\;\;\; \lambda_2=3(\omega_u-b),\;\;\;\;\;
\lambda_3=-3(b+\omega_x-\omega_u),\nonumber\\
&&\cdot \;\;\text{Point}\; B_2: \;\;\;\;\; \lambda_1=3b,\;\;\;\;\;
\lambda_2=-3\omega_x,\;\;\;\;\;
\lambda_3=3(b-\omega_u),\nonumber\\
&&\cdot \;\;\text{Point}\; C_2: \;\;\;\;\;
\lambda_1=3\omega_x,\;\;\;\;\; \lambda_2=3(\omega_x+b),\;\;\;\;\;
\lambda_3=3(b+\omega_x-\omega_u).
\end{eqnarray}
For point $A_2$, since $\omega_u>0$, the eigenvalue $\lambda_1$
always is positive, which indicates that $A_2$ is an unstable point.
Moreover, the EoS of DE $\omega_x<0$ and the coupling constant $b>0$
means that both of the eigenvalues $\lambda_1$ and $\lambda_2$ of
the point $B_2$ are positive. Thus the point $B_2$ is an unstable
point. For the point $C_2$,when $-\omega_x<b<\omega_u-\omega_x$, the
sign of $\lambda_2$ is always opposite to the signs of $\lambda_1$
and $\lambda_2$, which leads $C_2$ to a saddle point. However, when
$0<b<-\omega_x$, we find that all of eigenvalues ($\lambda_1$,
$\lambda_2$ and $\lambda_3$) are negative, which indicates that
$C_2$ is a stable point as shown in figure (3). Meanwhile, we also
find none of the coordinate components of $C_2$ in the phase space
disappears, which means that in this case there exist three
components (DE, DM and unparticle) in the late-time universe and the
coincidence problem can be alleviated. From Eq.(\ref{st1}), we learn
that the effective total EoS at point $C_2$ is
$\omega_{tot}=b+\omega_x$. Therefore we obtain
$\ddot{a}\propto-(1+3b+3\omega_{x})\;t^{\frac{2}{3(1+b+\omega_{x})}-2}$
and $\rho\propto a^{-3(1+b+\omega_{x})}$. It means that point $C_2$
is an accelerated scaling solution as $b<-\omega_x-1/3$ and there
exists singularity in the finite future as $b<-\omega_x-1$. For
fixed $\omega_x$, the effective total EoS increases with $b$. Figure
(\ref{2f}) shows that in the Universe described by $C_2$ the density
of the unparticle increases and DE decreases with the coupling
constant $b$, while the density of DM first increases and then
decreases. This means that the Universe will be dominated by
unparticle if the coupling constant $b$ is large enough.
\begin{figure}[ht]
\begin{center}
\includegraphics[width=6cm]{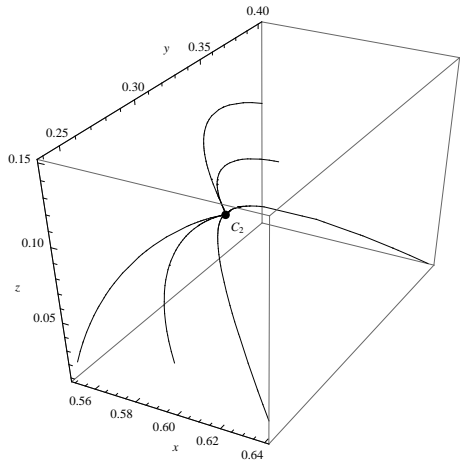}
\caption{The phase diagram of interacting dark energy with DM and
unparticle through the coupling terms (\ref{cou2}). The point $C_2$
is the critical point. Here we choose the values $\omega_x=-1.2$,
$\omega_u=0.28$ and $b=0.5$ in the stable region $b<-\omega_x$.}
\end{center}
\end{figure}
\begin{figure}[ht]
\begin{center}
\includegraphics[width=6cm]{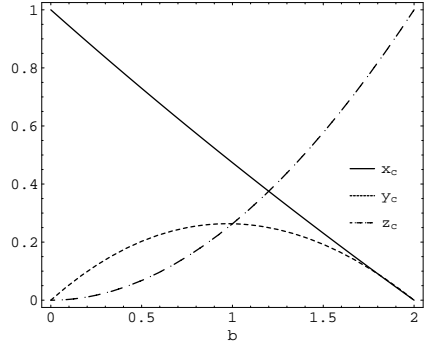}
\caption{Variety of $x_c$, $y_c$ and $z_c$ with $b$ at the critical
point $C_2$ for fixed $\omega_x=-1.8$ and $\omega_u=0.2$.}
\label{2f}
\end{center}
\end{figure}

\subsection{Interacting model III }

In Ref.\cite{sb1}, we have investigated the dynamics of the
interacting DE models in which the coupling terms $\Gamma_i$
contains the product of the densities of DE and DM and find that
this new type of dark sector coupling leads to the more interesting
accelerated scaling solutions and presents us more complicated
features in dynamical phase space. Therefore, in the following
sections, we will consider the cases in which the coupling terms
$\Gamma_i$ contain the product of the density of DE, DM and
unparticle and to see whether it presents some new properties or not
in the evolution of the Universe with the unparticle component. For
mathematical simplicity, we assume firstly the interaction among DE,
DM and unparticle have the forms
\begin{eqnarray}
\Gamma_1=-6b\kappa H^{-1}\rho_x\rho_u,\;\;\;\;\;
\Gamma_2=\Gamma_3=3b\kappa H^{-1}\rho_x\rho_u. \label{cou3}
\end{eqnarray}
As in the model (\ref{cou1}), these coupling terms also denote that
DE can be transfer to DM and unparticle. The dynamical equations of
the system can be written as
\begin{eqnarray}
x'&=&-6bxz-3\omega_xx+3x(\omega_xx + \omega_uz),\nonumber\\
y'&=&3bxz+3y(\omega_xx + \omega_uz),\nonumber\\
z'&=&3bxz-3\omega_uz+3z(\omega_xx + \omega_uz),
\end{eqnarray}
and the critical points $(x_c,\; y_c,\;z_c)$ are
\begin{eqnarray}
&&\cdot \;\;\text{Point}\; A_3: \;( 1,\;\;0,\;\;0),\nonumber \\ &&
\cdot \;\;\text{Point}\; B_3: \; ( 0,\;\;0,\;\;1),\nonumber \\
&&\cdot \;\;\text{Point}\; C_3: \; \bigg(
\frac{\omega_u(2b+\omega_x-\omega_u)}{b\;(2b+2\omega_x-\omega_u)},\;\;
\frac{(\omega_x-\omega_u)^2+3b(\omega_x-\omega_u)+2b^2}{b\;(2b+2\omega_x-\omega_u)},\;\;
-\frac{b(b+\omega_x-\omega_u)}{b\;(2b+2\omega_x-\omega_u)}\bigg).
\end{eqnarray}
The condition that the point $C_3$ exists is $b>\omega_u-\omega_x$.
Repeating the previous operations, we can obtain the $3\times 3$
matrix $\Xi$ of the linearized perturbation equations
\begin{eqnarray}
\Xi=\bigg[ \begin{array}{ccc} \omega_x(6x_c-3)+3\omega_uz_c-6b\;\;&0&3\omega_ux_c-6b \\
3(\omega_xy_c+bz_c) &3(\omega_xx_c+\omega_uz_c)& 3(\omega_uy_c+bx_c) \\
3(\omega_x+b)z_c &0 & 3\omega_x x_c+\omega_u(6z_c-3)+3b
\end{array}\bigg],
\end{eqnarray}
and the eigenvalues of the coefficient matrix for these critical
points
\begin{eqnarray}
&&\cdot \;\;\text{Point} \;A_3: \;\;\;\;\;
\lambda_1=3\omega_x,\;\;\;\;\; \lambda_2=3\omega_x,\;\;\;\;\;
\lambda_3=3(b+\omega_x-\omega_u),\nonumber\\
&&\cdot \;\;\text{Point}\; B_3: \;\;\;\;\;
\lambda_1=3\omega_u,\;\;\;\;\; \lambda_2=3\omega_u,\;\;\;\;\;
\lambda_3=-3(2b+\omega_x-\omega_u),\nonumber\\
&&\cdot \;\;\text{Point}\; C_3: \;\;\;\;\;
\lambda_1=\frac{3\omega_u\omega_x}{2b+2\omega_x-\omega_u},\;\;\;\;\;\nonumber\\
&&\lambda_{2,3}=\frac{3(b\omega_u\omega_x\pm\sqrt{(b\omega_u\omega_x[16b^3+8b^2(5\omega_x-4\omega_u)+4(\omega_x-\omega_u)^2(2\omega_x-\omega_u)+
b(32\omega^2_x-51\omega_x\omega_u+20\omega^2_u)]})}{2b(2b+2\omega_x-\omega_u)}.\nonumber\\
\end{eqnarray}
\begin{figure}[ht]
\begin{center}
\includegraphics[width=6cm]{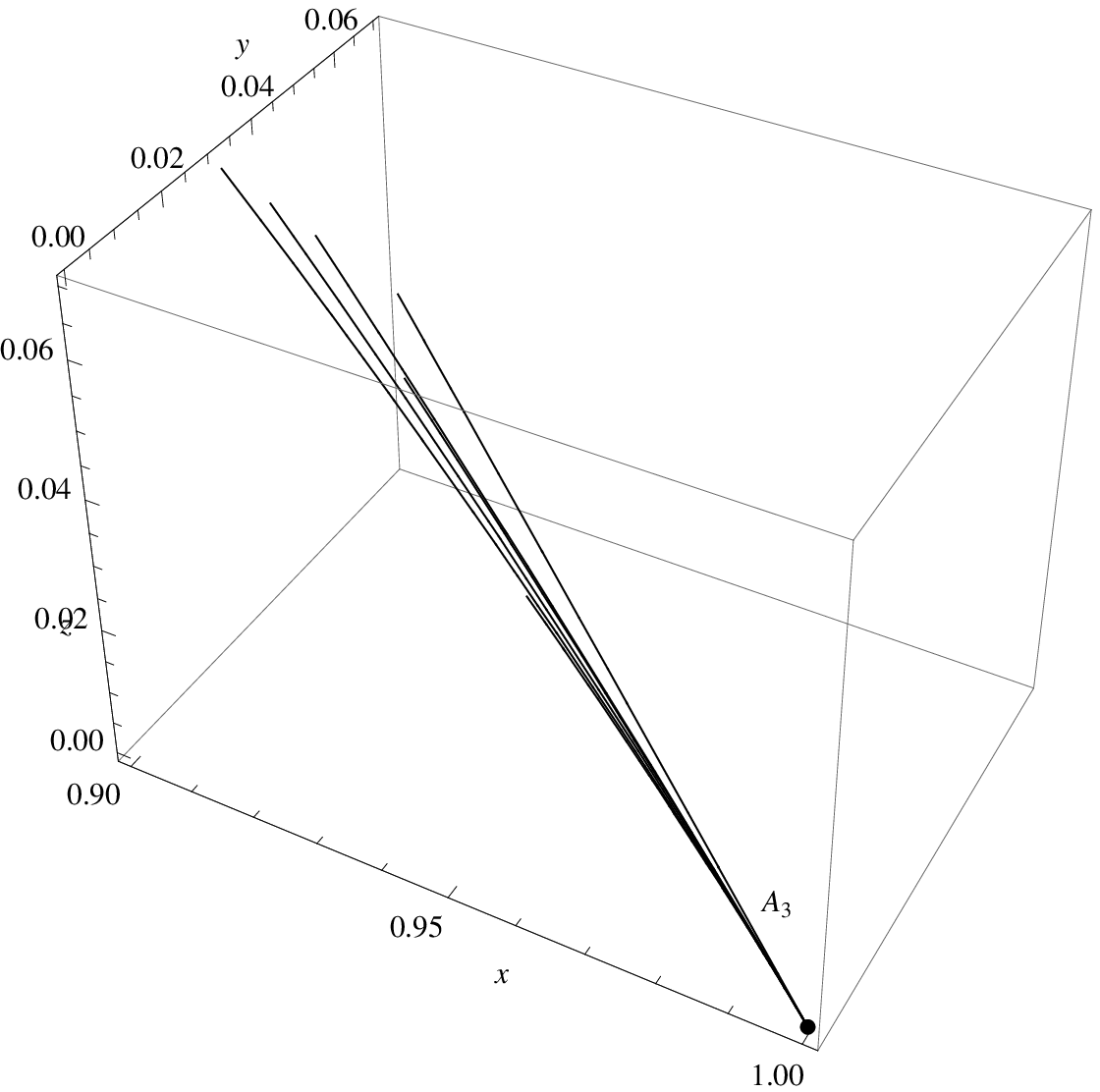}\;\;\;\;\;\includegraphics[width=6cm]{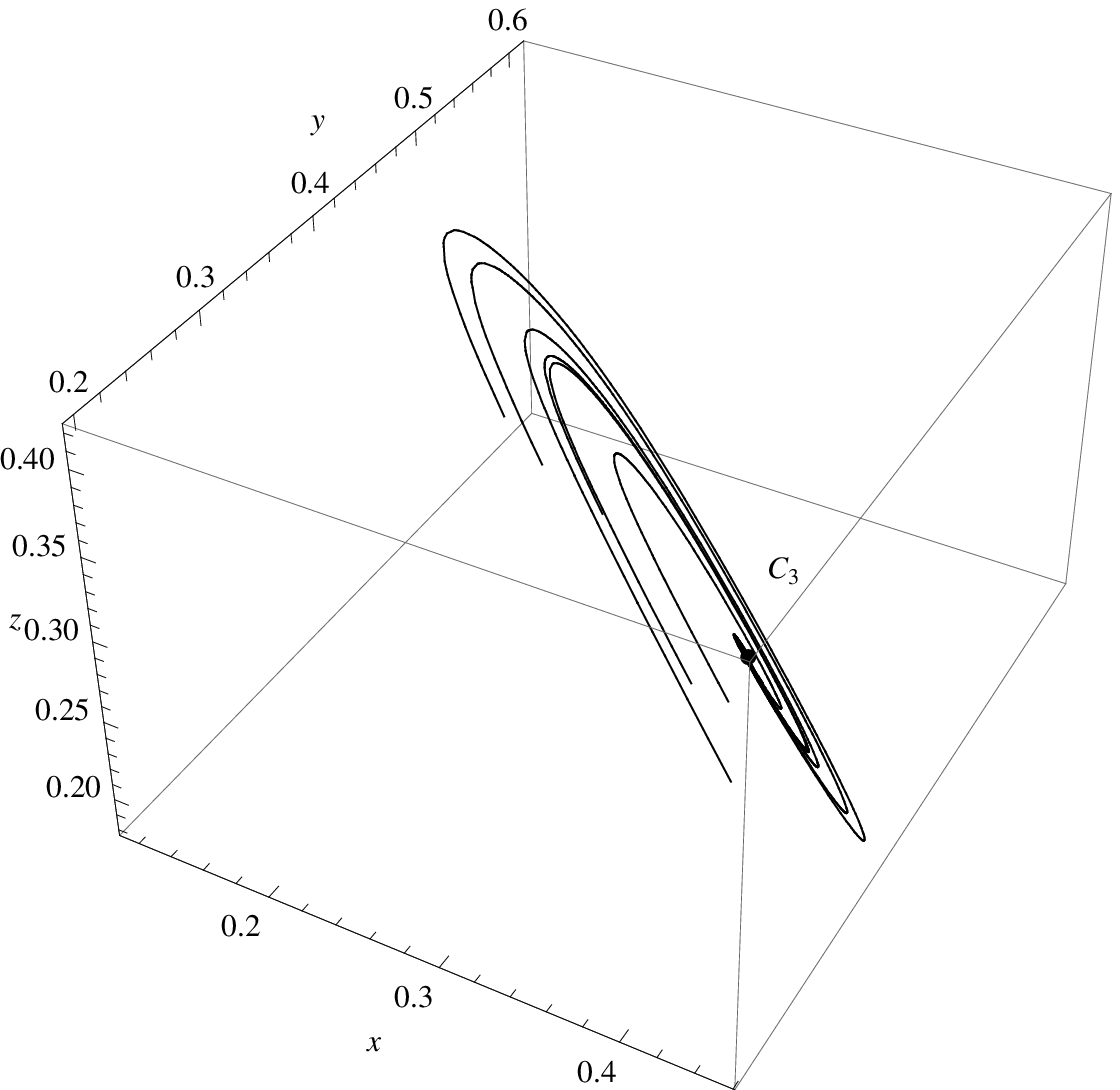}
\caption{The phase diagram of interacting dark energy with DM and
unparticle through the coupling terms (\ref{cou3}). In the left
figure, we fix the values ($\omega_x=-1.2$, $\omega_u=0.28$ and
$b=0.5$) which meets $b<\omega_u-\omega_x$ and the critical point is
$A_3$. In the right, we choose the values ($\omega_x=-1.2$,
$\omega_u=0.28$ and $b=1.8$) which satisfies $b>\omega_u-\omega_x$
and the critical point is $C_3$.}
\end{center}
\end{figure}
Through the similar analysis, we find that $A_3$ is a stable point
when $b<\omega_u-\omega_x$ and is a saddle point when
$b>\omega_u-\omega_x$. The point $B_3$ is an unstable point because
that $\omega_u>0$, which leads to the signs of the eigenvalues
$\lambda_{1,2}$ always are positive. For the point $C_3$, it is a
stable point when $b>\omega_x-\omega_u$. We can obtain the total
effective EoS
$\omega_{tot}=\frac{\omega_u\omega_x}{2b+2\omega_x-\omega_u}$. It
depends not only on the EoS of DE and unparticle, but also on the
coupling constant $b$. From the total effective EoS $\omega_{tot}$,
it is easy to obtain that point $C_3$ is an accelerated scaling
solution as $\omega_u-\omega_x<b
<(\omega_u-2\omega_x-3\omega_x\omega_u)/2$. While for
$b>(\omega_u-2\omega_x-3\omega_x\omega_u)/2$, we find that
$\omega_{tot}>-1/3$, which means that it is a decelerated scaling
solution. This can be explained by that the larger $b$ denote the
more DE transfers to DM and unparticle in the Universe. While, as
$b<\omega_x-\omega_u$, the point $C_3$ is an unstable point.
Moreover, in this model, when $b$ is small the coupling the Universe
will enter the era dominated by DE because $A_3$ is a stable point
(as shown in the left figure in Fig.(5)). When $b$ is larger the
Universe will enter a stable stage described by point $C_3$ (as
shown in the right figure in Fig.(5)), which contains the DE, DM and
unparticle. Thus, when $b>\omega_u-\omega_x$ this model can resolve
the coincidence problem. Moreover, from the Fig.(\ref{3f}), we also
find with the increase of $b$ the density of DM increases in the
Universe, but both of DE and unparticle decrease. This implies that
the Universe will be dominated by DM when the interaction is very
strong. Figure (\ref{4f}) also tells us that with the increase of
the coupling constant $b$ the EoS $\omega_{tot}$ increases so that
the Universe can make a transition from an accelerating expansion
phase to a decelerating one. As $b$ tends to infinity the EoS
$\omega_{tot}$ approaches to zero, which is consistent with that of
DM.
\begin{figure}[ht]
\begin{center}
\includegraphics[width=6cm]{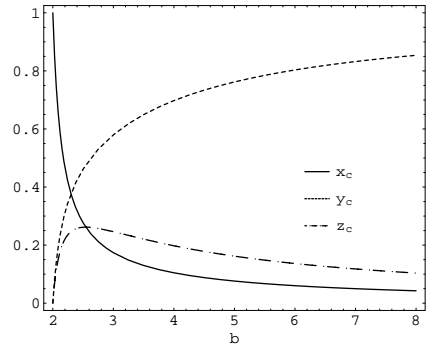}
\caption{Variety of $x_c$, $y_c$ and $z_c$ with $b$ at the critical
point $C_3$ for fixed $\omega_x=-1.7$ and $\omega_u=0.3$. Here the
coupling constant $b$ is located in the region
$b>\omega_u-\omega_x$.} \label{3f}
\end{center}
\end{figure}
\begin{figure}[ht]
\begin{center}
\includegraphics[width=6cm]{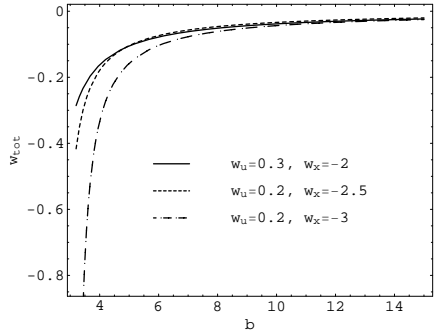}
\caption{Variety of the effective EoS $\omega_{tot}$ with $b$ at the
critical point $C_3$ for different $\omega_x$ and $\omega_u$. Here
the coupling constant $b$ is limited in the region
$b>\omega_u-\omega_x$.} \label{4f}
\end{center}
\end{figure}

\subsection{Interacting model IV }

In the model (III), we consider only the case the coupling terms do
not contain the density of DM. In the following model, we will
consider the interaction includes $\rho_m$ and suppose that
$\Gamma_i$ have the forms
\begin{eqnarray}
\Gamma_1=-3b\kappa H^{-1}\rho_x\rho_u,\;\;\;\;\; \Gamma_2=3b\kappa
H^{-1}(\rho_x\rho_u-\rho_m\rho_u),\;\;\;\;\; \Gamma_3=3b\kappa
H^{-1}\rho_m\rho_u. \label{cou4}
\end{eqnarray}
Although the coupling forms are more complicated than the model
(\ref{cou3}), it also presents us the analytical expressions for the
critical points which brings us the simplicity in the later
calculation. This type of $\Gamma_i$ also denote DE can be transfer
to DM and at the same time DM can also be convert to unparticle. The
dynamical equations of the system are given by
\begin{eqnarray}
x'&=&-3bxz-3\omega_xx+3x(\omega_xx + \omega_uz),\nonumber\\
y'&=&3b(xz-yz)+3y(\omega_xx + \omega_uz),\nonumber\\
z'&=&3byz-3\omega_uz+3z(\omega_xx + \omega_uz),
\end{eqnarray}
and the critical points $(x_c,\; y_c,\;z_c)$ can be expressed as
\begin{eqnarray}
&&\cdot \;\;\text{Point}\; A_4: \;( 1,\;\;0,\;\;0),\nonumber \\ &&
\cdot \;\;\text{Point}\; B_4: \; ( 0,\;\;0,\;\;1),\nonumber \\
&&\cdot \;\;\text{Point}\; C_4: \; \bigg(
\frac{b+\omega_x-\omega_u}{b},\;\;
\frac{(\omega_u-\omega_x)(b+\omega_x-\omega_u)}{b\;(b-\omega_u)},\;\;
-\frac{\omega_x(\omega_x-\omega_u)}{b\;(b-\omega_u)}\bigg).
\end{eqnarray}
The critical point $C_4$ exists only when $b>\omega_u-\omega_x$.
Similarly, we can obtain the $3\times 3$ matrix $\Xi$
\begin{eqnarray}
\Xi=\bigg[ \begin{array}{ccc} \omega_x(6x_c-3)+3\omega_uz_c-3b\;\;&0&3(\omega_u-b)x_c \\
3(\omega_xy_c+bz_c) &3(\omega_xx_c+\omega_uz_c-bz_c)& 3(\omega_uy_c+bx_c-by_c) \\
3\omega_xz_c &3bz_c & 3\omega_x x_c+\omega_u(6z_c-3)+3by_c
\end{array}\bigg].
\end{eqnarray}
The eigenvalues of the coefficient matrix for these critical points
are
\begin{eqnarray}
&&\cdot \;\;\text{Point} \;A_4: \;\;\;\;\;
\lambda_1=3\omega_x,\;\;\;\;\; \lambda_2=3\omega_x,\;\;\;\;\;
\lambda_3=3(\omega_x-\omega_u),\nonumber\\
&&\cdot \;\;\text{Point}\; B_4: \;\;\;\;\;
\lambda_1=3\omega_u,\;\;\;\;\; \lambda_2=3(\omega_u-b),\;\;\;\;\;
\lambda_3=-3(b+\omega_x-\omega_u),\nonumber\\
&&\cdot \;\;\text{Point}\; C_4: \;\;\;\;\;
\lambda_1=\frac{3\omega_x(b+\omega_x-2\omega_u)}{b-\omega_u},\;\;\;
\lambda_{2,3}=\frac{3(b\omega_x\pm\sqrt{b\omega_x[4(\omega_x-\omega_u)^2+b(5\omega_x-4\omega_u)]
})}{2b}.
\end{eqnarray}
Obviously, all of eigenvalues ($\lambda_1$, $\lambda_2$ and
$\lambda_3$) for the point $A_4$ are negative because the EoS of DE
$\omega_x<0$, which indicates that in this case the point $A_4$
always is a stable point. The point $B_4$ is an unstable point since
$\lambda_1$ is positive. From the previous discussion we find that
the point $C_4$ exists only when $b>\omega_u-\omega_x$, which leads
to that $\lambda_2>0$ and $\lambda_3<0$. This means that the point
$C_4$ is a saddle point. Since the point $A_4$ describes a Universe
filled with only DE, the interaction (\ref{cou4}) cannot resolve the
coincidence problem.

\section{Conclusions and discussions}

In this paper we have studied the dynamical behaviors when DE is
coupling to DM  and unparticle in the standard flat FRW cosmology.
We considered four different interacting models and examined the
stability of critical points. In all the examined models we found
that there exist a stable late-time scaling solution which
corresponds to an accelerating universe. This feature was expected
since DE cosmology has been constructed in order to always satisfy
this condition. Moreover, for all the studied cases, we also find
the accelerating stable solutions if we choose the appropriated
parameters in the models. Except the fourth model, the stable
solutions in other models admit that DE coexists with DM and
unparticle in the Universe. Our result also implies that in these
models the coincidence problem can be alleviated only when the
coupling is not strong enough.

\begin{acknowledgments}
This work was partially supported by the National Natural Science
Foundation of China under Grant No.10875041; the Scientific Research
Fund of Hunan Provincial Education Department Grant No.07B043 and
the construct program of key disciplines in Hunan Province. J. L.
Jing's work was partially supported by the National Natural Science
Foundation of China under Grant No.10675045 and No.10875040; and the
Hunan Provincial Natural Science Foundation of China under Grant
No.08JJ3010.
\end{acknowledgments}

\end{document}